\documentclass[a4]{spie}  %>>> use for US letter paper
 \usepackage[dvipsnames]{xcolor}
\usepackage{amsmath,amsfonts,amssymb}
\usepackage{graphicx}
\usepackage[colorlinks=true, allcolors=blue]{hyperref}
\usepackage{multirow}
\newcommand{\figref}[1]{\figurename~\ref{#1}}
\newcommand{\tabref}[1]{\tablename~\ref{#1}}
\usepackage{xspace}
\newcommand{\londonscrc}{KCLH MC-RC\xspace}
\newcommand\tab[1][1cm]{\hspace*{#1}}
\title{A Clinical Guideline Driven Automated Linear Feature Extraction for Vestibular Schwannoma}

\author[a]{Navodini Wijethilake}
\author[a,b]{Steve Connor}
\author[c]{Anna Oviedova}
\author[b]{Rebecca Burger}
\author[a]{Tom Vercauteren}
\author[a,c]{Jonathan Shapey}
\affil[a]{School of BMEIS, King’s College London, London, United Kingdom}
\affil[b]{Department of Neuroradiology, King’s College Hospital, London, United Kingdom}
\affil[c]{Department of Neurosurgery, King’s College Hospital, London, United Kingdom}

\authorinfo{Further author information: (Send correspondence to N.W.)\\E-mail: navodini.wijethilake@kcl.ac.uk}

\begin{document} 
\maketitle

\begin{abstract}
Vestibular Schwannoma is a benign brain tumour that grows from one of the balance nerves. Patients may be treated by surgery, radiosurgery or with a conservative "wait-and-scan" strategy. Clinicians typically use manually extracted linear measurements to aid clinical decision making. This work aims to automate and improve this process by using deep learning based segmentation to extract relevant clinical features through computational algorithms. To the best of our knowledge, our study is the first to propose an automated approach to replicate local clinical guidelines. Our deep learning based segmentation provided Dice-scores of 0.8124 $\pm$ 0.2343 and 0.8969 $\pm$ 0.0521 for extrameatal and whole tumour regions respectively for T2 weighted MRI, whereas 0.8222 $\pm$ 0.2108 and 0.9049 $\pm$ 0.0646 were obtained for T1 weighted MRI. We propose a novel algorithm to choose and extract the most appropriate maximum linear measurement from the segmented regions based on the size of the extrameatal portion of the tumour. Using this tool, clinicians will be provided with a visual guide and related metrics relating to tumour progression that will function as a clinical decision aid. In this study, we utilize 187 scans obtained from 50 patients referred to a tertiary specialist neurosurgical service in the United Kingdom. The measurements extracted manually by an expert neuroradiologist indicated a significant correlation with the automated measurements ($p < 0.0001$).
\end{abstract}

% Include a list of keywords after the abstract 

% \section{INTRODUCTION}
% \label{sec:intro}  % \label{} allows reference to this section
\section{Introduction}
\begin{figure}[tb!]
\includegraphics[width=\textwidth]{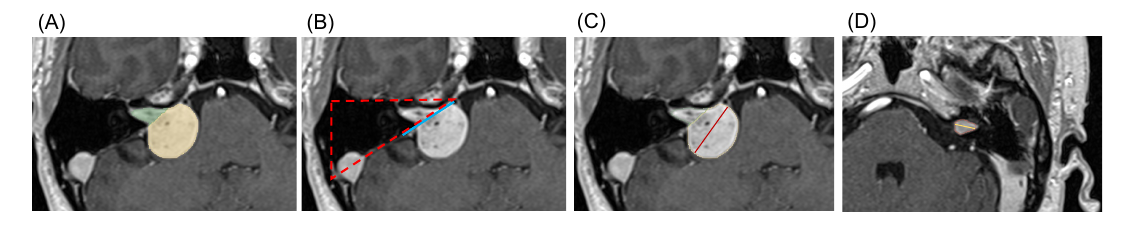}
\caption{\textbf{A:} intra-/extra-meatal regions are shown in green and yellow labels respectively. \textbf{B:} Petrous pyramid region is shown in red dashed margin and the boundary separating intra-/extra-meatal regions is shown in blue. \textbf{C:} Maximum extrameatal diameter. \textbf{D:} Maximum whole tumour diameter. } \label{fig1:anatomy}
\end{figure}

Vestibular Schwannoma (VS) is the most common adult brain tumour that grows in the cerebellopontine angle, accounting for 8\% of all the intracranial brain tumours. Patients may be treated by surgery, radiosurgery or with a conservative "wait-and-scan" strategy. Irrespective of the type of treatment received, most patients require prolonged radiological follow up and regular surveillance imaging \cite{shapey2018standardised}. The choice and timing of treatment depends on the patient's symptoms and the behaviour of the tumour \cite{prasad2018decision}. For this a consistent and a standardized measurement of the tumour should be extracted and reported. The Consensus Meeting on Systems for Reporting Results in Acoustic Neuroma (2001) advise that: 1) The intrameatal and extrameatal (see \figref{fig1:anatomy}.A) portion of the VS should be clearly distinguished; and 2) the largest extrameatal diameter (see \figref{fig1:anatomy}.C) should be measured
\cite{kanzaki2003new}.
If the tumour only has an intrameatal portion, the maximum tumour diameter of the whole tumour region (which is also the intrameatal portion) is presented according to the expert neuroradiologist (see \figref{fig1:anatomy}.D).

In current clinical settings, these features are extracted manually by a neuroradiologist, prior to a multi-disciplinary meeting where treatment decisions are made. This manual feature extraction task is typically performed by an experienced neuroradiologist but is time-consuming, labour-intensive  and prone to variability and subjectivity.  

In the past decade, the field of radiomics has shown the potential to address challenges related to diagnostic and prognostic tasks in vari
ous pathologies. Automated feature extraction platforms such as pyradiomics \cite{pyradiomics} and LIFEx \cite{nioche2018lifex} have in particular matured to comply with the IBSI (Image Biomarkers Standardization Initiative) standards \cite{hatt2018ibsi}. However, in neuroimaging, these efforts have largely been focused on malignant brain tumours. There thus remains a gap between existing research platforms and clinical expectations for VS management. For example, VS tumour diameter on an axial MR slice should not only be extracted but should also be visualised in context. Existing platforms focus on measurement extraction that may not be tailored for VS and importantly provide no means of visualising these. 

Despite the prevalence of linear measurements in current practice,
volumetric tumour measurements have been shown to be the most accurate and sensitive measure of detecting VS growth \cite{mackeith2018comparison,shapey2021segmentation}. Implementing routine volumetric measurements would enable clinicians to offer earlier appropriate intervention.
However, currently available tools make calculating VS tumour volume assessment a labour-intensive process, prone to variability and subjectivity, with no dedicated software readily available within the clinical setting. Artificial Intelligence (AI)-driven clinical support tools have the potential to improve patient outcomes and experience by the standardisation and personalisation of VS treatment \cite{shapey2021artificial}.
%In our previous work, we have generated a AI framework 
Previous research work has indeed demonstrated that AI tools are technically capable of automatically detecting and segmenting VS
\cite{aaronnmm,shapey2019artificial} and even delineating intra-/extra-meatal components \cite{wijethilake2022boundary}. Such tools can serve as a foundation for the automated feature extraction task. 

In this work, we propose an automated clinical feature extraction framework for VS. First, we develop deep learning based segmentation models for intra-/extra-meatal segmentation for different MRI modalities. Using the tumour contours, we then develop feature extraction algorithms to measure the tumour size while presenting the most appropriate linear measurement based the tumour segmentation. We analyse the relationship between our extracted features and the manually extracted features by an expert neuroradiology consultant(S.C.) for 50 patients. 

\section{Material and Methods}

\begin{figure}[t!]\centering
\includegraphics[width=0.99\textwidth]{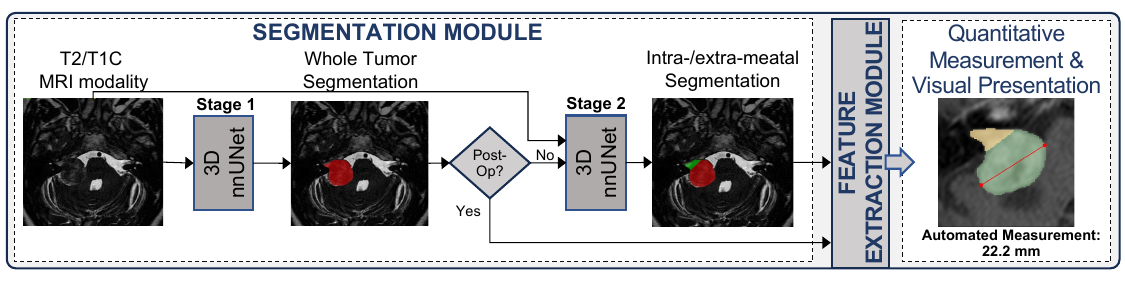}
\caption{The proposed framework: Segmentation module consists of a two-stage approach for pre-operative scans, whereas the finetuned stage 1 model is used for the post-operative scans. This is integrated with the feature extraction module to provide clinical features for 50 patients.} \label{fig1:outline}
\end{figure}

\subsection{Datasets for Deep Learning based Segmentation}
We build on the UCLH MC-RC (Multi-centre Routine-Clinical) dataset \cite{aaronnmm} for the segmentation task. The UCLH MC-RC dataset comprised MRI scans of 165 patients acquired in the period between February 2006 and September 2019. All the patients were over 18 years old and diagnosed with unilateral VS. Due to the low availability of the contrast enhanced T1 modality (T1C) in the UCLH MC-RC dataset, the London SC-GK (Single-centre Gamma-knife) \cite{shapey2021segmentation} and the Tilburg SC-GK datasets \cite{dorent2023crossmoda} were combined with the UCLH MC-RC dataset. Each SC-GK dataset consists of T1C and T2 for patients undergoing Gamma Knife Stereotactic Radiosurgery (GK SRS) from two institutes in London, UK and Tilburg, Netherlands. In the two-stage segmentation pipeline illustrated in \figref{fig1:outline} both stages were initially trained with pre-operative scans. Stage 1 was subsequently fine-tuned with the post-operative scans from UK MC-RC and London SC-GK datasets. 

The models for pre-operative intra-/extra-meatal segmentation and for post-operative whole tumour segmentation were developed using T1C and T2 modalities separately. \tabref{tab1:dataset_distribution} summarizes the number of patients ($N_p$) and number of scans ($N_s$) in each training, validation and testing set from three utilized datasets.  

\subsection{\londonscrc Dataset for Feature Extraction}
The dataset for feature extraction was obtained retrospectively from King's College Hospital, London UK. All patients had previously been managed under the care of the multidisciplinary skull base team and first presented to the service between December 2009 and March 2012. Patients with a single unilateral VS, aged over 18 years were eligible for inclusion in the dataset. Patients with Neurofibromatosis type 2 (NF2) were excluded. Each patient had imaging and clinical data from at least 1 timepoint. Available MR imaging included pre- and post-contrast T1 imaging and/or T2 weighted imaging. MRI had a maximum slice thickness of 3.5\,mm. 
50 patient cases with available MRI and clinical data for a minimum of 1 time point (range 1 - 6 consecutive times points) were selected for this study. The duration of follow up ranged from 1 to 11 years. If a patient session included both T1C and T2 MRI data, the T2 data was used to generate the segmentation. In total, 187 sessions across 50 patients comprising 145 T1C scans and 42 T2 scans respectively were used for the feature extraction. 23 sessions out of 187 were post operative scans. 

This study was approved by the
\ifdefined\showidentifiedinfo
NHS Health Research Authority and Research Ethics Committee (18/LO/0532).
\else
relevant authorities (anonymised).
\fi
Because patients were selected retrospectively and the MR images were de-identified before analysis, no informed consent was required for the study. 

\begin{table}[tb!]
\caption{Distribution of data between training, validation and testing sets across the three datasets used for segmentation task.}\label{tab1:dataset_distribution}
\centering
\begin{tabular}{|l|l|l|l|l|l|l|l|l|l|l|l|l|l|}
\hline
\multirow{3}{*}{} &  \multirow{3}{*}{Dataset} & \multicolumn{6}{|l|}{T1C} & \multicolumn{6}{|l|}{T2}\\
\cline{3-14}
 &   & \multicolumn{2}{|l|}{Train} & \multicolumn{2}{|l|}{Valid}  & \multicolumn{2}{|l|}{Test} & \multicolumn{2}{|l|}{Train} & \multicolumn{2}{|l|}{Valid} & \multicolumn{2}{|l|}{Test}\\
\cline{3-14}

 &   & $N_{p}$ & $N_{s}$ & $N_{p}$ & $N_{s}$ & $N_{p}$ & $N_{s}$ & $N_{p}$ & $N_{s}$& $N_{p}$ & $N_{s}$& $N_{p}$ & $N_{s}$\\
\hline
\multirow{3}{*}{Pre-Op} & UK MC-RC & 41 & 53 & 6 & 8 & 12 & 15 &97 & 212& 17 & 35 & 30 & 61 \\ \cline{2-14}
 & London SC-GK & - & 55 & - & 9 &-  &15 & -&- &- & -&  -& -\\  \cline{2-14}
 & Tilburg SC-GK & - & 72 & - & 12 & -  &21 &- &- &- &- & - & -\\  \hline
\multirow{3}{*}{Post-Op} & UK MC-RC& 4& 5 &0 & 0&1 & 1   & 20 & 33 &3 & 5 &  5& 10\\ \cline{2-14}
 & London SC-GK & -& 52 & -& 8 & - & 14 & -& 52 & -& 8 & - & 14\\  \hline
\end{tabular}
\end{table}

\subsection{Deep Learning based Segmentation}
%\subsection{Training}
Following the approach of \cite{wijethilake2022boundary}, we make use of the default 3D full resolution UNet of the nnU-Net framework (3D nnU-Net) \cite{isensee2021nnu} in two stages sequentially, to obtain the whole tumour and intra-/extra-meatal segmentation in each stage respectively. Whole tumour masks for training the stage 2 are generated by performing 5-fold cross validation in the stage 1. A loss combining Cross-entropy and Dice score is used in both stages. 
The pre-trained model for the stage 1 (whole tumour segmentation task) with pre-operative sessions, was finetuned on post-operative scans independently. 

%\subsection{Validation}
The results were validated and tested on the corresponding validation and testing sets. 
The best performing models were used to obtain the segmentations of the \londonscrc dataset. Consequently, the 2-stage nnUNet trained for split segmentation of the pre-operative scans or fine-tuned model for whole tumour segmentation of the post-operative scans were utilized based on the pre-/post-operative condition of each session in the \londonscrc dataset.

\subsection{Automated Feature Extraction Overview}
For the \londonscrc dataset, the obtained segmentation masks were used for the feature extraction task. Largest component analysis was performed prior to the feature extraction to exclude the false positives not connected to the tumour component.

We extracted the maximum tumour diameter as the key linear measurement for VS. This measurement could be extracted from the whole tumour region ($\mathcal{D}^{WT}$) and/or from the extrameatal region of the tumour ($\mathcal{D}^{EM}$). If the extrameatal tumour portion is small/negligible, radiologists typically obtain the linear measurement for the the whole tumour region (both intra-/extra-meatal regions combined).

\subsubsection{Maximum Tumour / Tumour Region Diameter ($\mathcal{D}$)} 
Let $\mathcal{S}_x$ be the tumour mask of patient $x$, generated using the DL model. We obtain the boundary of the region of interest (ROI) ($\mathcal{B}_{x}$) using sklearn \texttt{find\_boundaries} function.
%, based on the morphological operations; dilation and erosion.  
%
The upper and lower convex hulls ($U_{x}$ and $L_{x}$) of the $\mathcal{B}_{x}$ structure is obtained using the Andrew's monotone chain algorithm, a modified version of Graham's scan algorithm \cite{andrew1979another}. Afterwards, rotating caliper method is used to obtain the maximum tumour diameter ($\mathcal{D}$). 

This feature $\mathcal{D}$ can be extracted either from whole tumour ($\mathcal{D}^{WT}$) or from the extrameatal region ($\mathcal{D}^{EM}$).
To reduce the cognitive burden on the clinical team, only a single of these diameters is extracted and displayed as discussed next.

\subsubsection{Selection of the Most Relevant Diameter to Display}

The decision of which diameter to present to the users is conditioned on the operative status of the patient and three additional features:
1) maximum intrameatal diameter parallel to the posterior petrous pyramid ($\textbf{d}_{(intra,\parallel)}$);
2) maximum extrameatal diameter parallel to the posterior petrous pyramid ($\textbf{d}_{(extra,\parallel)}$);
and 3) maximum extrameatal diameter perpendicular to the posterior petrous pyramid ($\textbf{d}_{(extra,\bot)}$). For post-operative cases, the maximum whole tumour diameter is extracted.

For pre-operative cases, we consider two scenarios.
The first is when the predicted mask is completely intracanalicular, i.e. only an intrameatal region is present. Then, the intrameatal region is the whole tumour and we extract its maximum diameter. The second is when the predicted mask has an extrameatal portion.
We then look into three additional features before deciding which diameter should be extracted. 
In pseudo-code, the following decision rule is used:\\
    \textbf{IF} $\textbf{d}_{(intra,\parallel)}$ $\ge$ $\textbf{d}_{(extra,\parallel)}$ \\
    \tab $\rightarrow$ use $\mathcal{D}^{WT}$ \\
    \textbf{ELSE} \\
    \tab \textbf{IF} 
    $\textbf{d}_{(extra,\bot)}$ $> 2mm$ \\
    \tab \tab $\rightarrow$
    use $\mathcal{D}^{EM}$ \\
    \tab \textbf{ELSE IF} 
    $\textbf{d}_{(extra,\bot)}$ $\leq 2mm$ \\
    \tab \tab $\rightarrow$ 
    use $\mathcal{D}^{WT}$
    
\subsection{Manual Feature Extraction} 
The manual feature extraction was performed by an expert neuroradiologist
\ifdefined\showidentifiedinfo
(S.C.)
\fi
blinded to the automated segmentation masks and the automated measurements. Following current clinical practice for MDT meetings, three sessions (index, second most recent, most recent) from 50 patients were chosen for feature extraction. Thus, 140 sessions out of 187 were selected. This was performed on PACS workstation (Sectra workstation, Sectra AB, Sweden) and under time constrained conditions to mimic clinical MDM preparation. The neuroradiologist chose $\mathcal{D}^{EM}$ or $\mathcal{D}^{WT}$, based on whether the $\mathcal{D}^{WT}$ appeared larger than the porus on axial images. 

\subsection{Analysis of Agreement between Manual and Automated Linear Measurements}
Comparison of manual and automated measurements were performed including:
1) pre-operative extrameatal tumour measurements;
2) pre-operative whole tumour measurements (exclusively intrameatal tumours);
and 3) post-operative whole tumour measurements.
Within each category, the correlation between the two measurements were assessed with the Pearson Correlation. Further, Bland–Altman plots were generated to visualize the level of agreement between manual and the automated linear measurements. The mean measurement difference (Automated Measurement - Expert Manual Measurement) is presented as bias together with its standard deviation and 95\% agreement lines.

\section{Results}

\subsection{Deep Learning based Segmentation}
\tabref{tab1:segmentation_dataset} summarizes the performance of the segmentation models during the testing phase.
The segmentation masks of \londonscrc dataset were also quality evaluated by an expert neuroradiologist.
\ifdefined\showidentifiedinfo
(S.C.).
\fi 

\begin{table}[htb!]
\caption{Quantitative results of the segmentation task, for intra-/extra-meatal segmentation and whole tumour segmentation.}\label{tab1:segmentation_dataset}
\centering
\begin{tabular}{|l|l|l|l|l|}
\hline
MRI Modality &  Case & Intrameatal & Extrameatal & Whole Tumour \\ \hline
\multirow{2}{*}{T2} & Stage 1 (Pre-op)&	- & - & 0.8981$\pm$ 0.0528 \\ \cline{2-5}
& Stage 2 (Pre-op) & 0.7684$\pm$ 0.1065&	0.8124	$\pm$ 0.2343	& 0.8969$\pm$ 0.0521	\\ \cline{2-5}
& Fine tuning (Post-op)	& -  & - & 0.8191 $\pm$ 0.1891  \\ \hline

\multirow{2}{*}{T1} & Stage 1 (Pre-op)	& - & - & 0.9173	$\pm$ 0.0457 	 \\ \cline{2-5}
& Stage 2 (Pre-op) &  0.7927 $\pm$ 0.1446 &	0.8222$\pm$ 0.2108&	0.9049$\pm$ 0.0646 \\ \cline{2-5}
& Fine tuning (Post-op)	& - & - & 	0.8510 $\pm$ 0.0878  \\ \hline
\end{tabular}
\end{table}

\begin{figure}[htb!]
\centering
\includegraphics[width=0.6\textwidth]{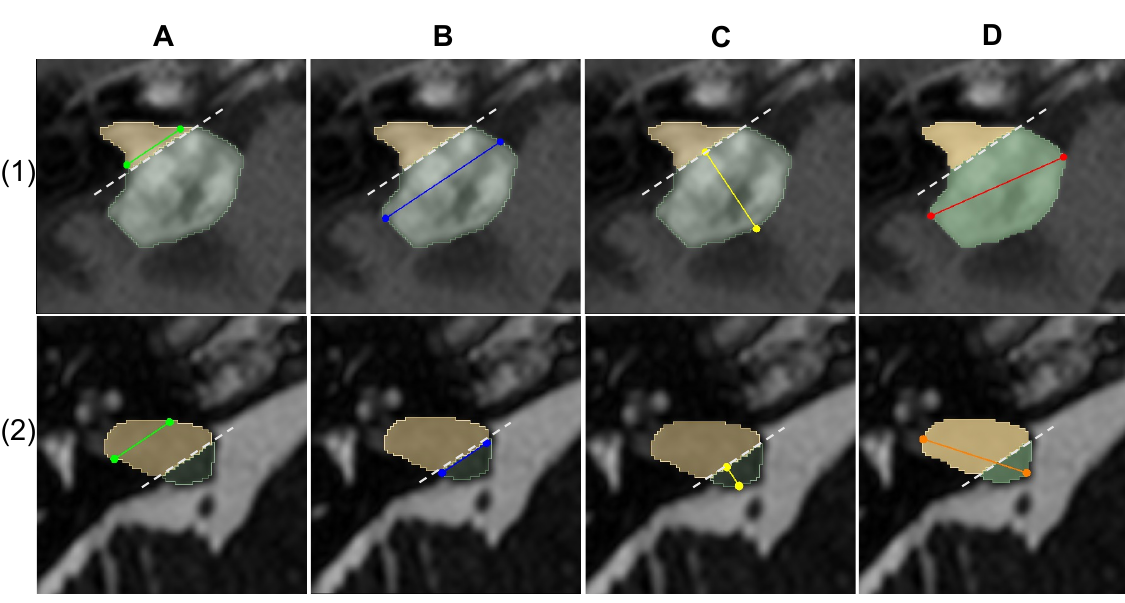}
\caption{Rows \textbf{1} \& \textbf{2} are different cases from the \londonscrc dataset. Columns \textbf{A}, \textbf{B}, \textbf{C} show the additional features measured when deciding the presented feature in the column \textbf{D}. The \textcolor{green}{green} line in column \textbf{A} is the $\textbf{d}_{(intra,\parallel)}$. The \textcolor{blue}{blue} line in column \textbf{B} is the $\textbf{d}_{(extra,\parallel)}$. The \textcolor{yellow}{yellow} line in column \textbf{C} is the $\textbf{d}_{(extra,\bot)}$. The posterior petrous pyramid, or the boundary between the intra-/extra-meatal region is shown in white dotted line. In the column \textbf{D}, the maximum extrameatal diameter is shown in \textcolor{red}{red} and the maximum whole tumour diameter is shown in \textcolor{orange}{orange}.} \label{fig1:results_featureextraction}
\end{figure}

\subsection{Feature Extraction}

\figref{fig1:results_featureextraction} shows 2 instances from the \londonscrc dataset, which were employed for feature extraction task. \figref{fig1:results_featureextraction}.1 shows an instance where both have fulfilled the criteria to select the maximum diameter of the extrameatal region is calculated ($\mathcal{D}^{EM}$). \figref{fig1:results_featureextraction}.2 shows an instance where this has fulfilled the criteria to present the maximum diameter of the whole tumour ($\mathcal{D}^{WT}$).

%\subsubsection{Comparison between Expert Manual Measurements and Automated Measurements}
\subsubsection{Extrameatal Tumour Measurement Comparison for Pre-operative Tumours with an Extrameatal Component}
70 out of 140 sessions were pre-operative with a large extrameatal portion. In these cases the expert clinical manually extracted the extrameatal diameter. Our segmentation algorithm failed to provide an acceptable segmentation for 3/70 instances.
In a further 2 instances, the clinician chose to provide a whole tumour measurement whereas the automated algorithm selected the extrameatal diameter instead. These cases were also excluded from further analysis. Consequently, we assessed the correlation and agreement between extrameatal diameter measurements on 65 scans. A significant positive correlation ($r=0.9909$,$p < 0.0001$) was found between the manual and automated measurements (\figref{fig1:comparison_featureextraction}.A) with a bias of $0.777\,mm\pm 1.057$ was observed in the difference between automated and the manual measurements (\figref{fig1:comparison_featureextraction}.D). 

\subsubsection{Whole Tumour Measurement Comparison for Pre-operative Intrameatal-only Tumours}
50 out of 140 sessions were pre-operative and intrameatal only, for which the clinician measured the whole tumour diameter. Our segmentation algorithm did not provide acceptable segmentation for 10 sessions out of 50. For 2 sessions our algorithm extracted the extrameatal diameter, whereas the clinician chose the whole tumour measurement. Thus, Altogether 12 instances were excluded from the analysis. Subsequently, we analysed the difference of the whole tumour diameter between the manual and automated measurements. In \figref{fig1:comparison_featureextraction}.B, we have shown the correlation between the two independent measurements ($r=0.8693$,$p < 0.0001$) and \figref{fig1:comparison_featureextraction}.E illustrates the Bland-Altman Plot. The results indicated a bias of $-0.069\,mm \pm 1.635$.

\subsubsection{Whole Tumour Measurement Comparison for Post-operative Tumours}
\begin{figure}[tbh!]
\centering
\includegraphics[width=0.89\textwidth]{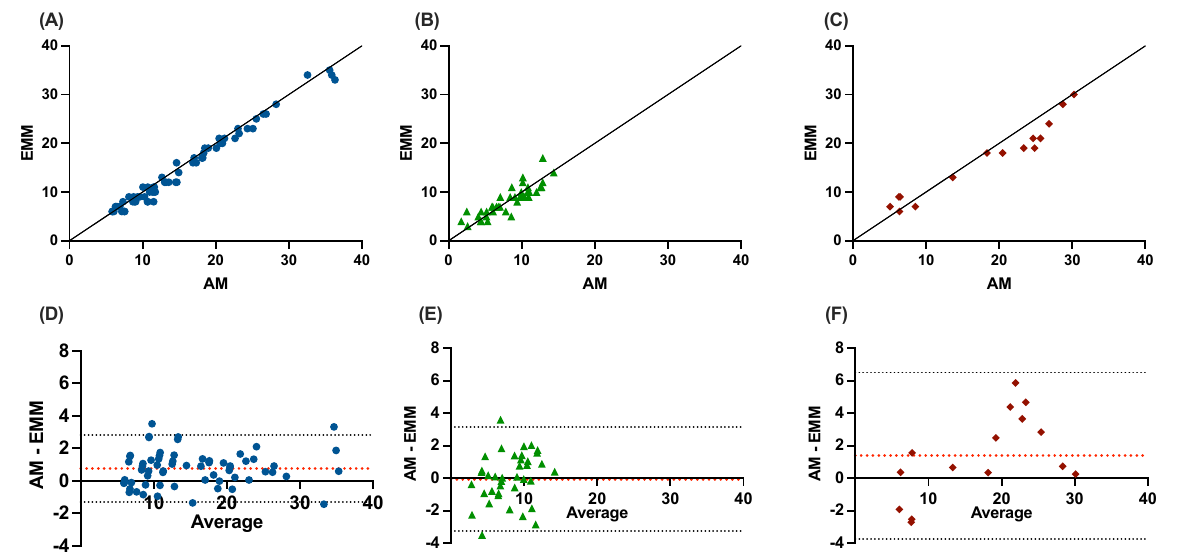}
\caption{Plots with the assessment of agreement between two methods of clinical measurement; automated vs manual. \textbf{A} \& \textbf{D:} represent the relationship between maximum extrameatal diameter feature. \textbf{B} \& \textbf{E:} represent the relationship between maximum whole tumour diameter feature for pre-operative sessions. \textbf{C} \& \textbf{F:} represent the relationship between maximum whole tumour diameter feature for post-operative sessions. In Bland-Altman plots, red dashed lines indicate the bias, (mean of the difference) and black dashed lines indicate upper and lower 95\% limits of agreement. \textbf{AM:} Automated Measurement, \textbf{EMM:} Expert Manual Measurement} \label{fig1:comparison_featureextraction}
\end{figure}

20 sessions were post-operative and the clinicians chose the maximum whole tumour diameter as the key measurement. The clinician chose not to extract features from 2 sessions as they were negligible in size and our segmentation model provided an under-segmented mask for 3 sessions. We therefore analysed 15 sessions with post-operative imaging. A significant correlation was noted ($r=0.9689$,$p < 0.0001$) (\figref{fig1:comparison_featureextraction}.C), while Bland-Altman analysis showed a bias of $1.369\,mm \pm 2.614 $ in the difference between automated and manual measurements (\figref{fig1:comparison_featureextraction}.F). 

\section{Discussion and Conclusion}
Our study propose an automated feature extraction framework for Vestibular Schwannoma management. In this work, we utilized deep learning based segmentation along with computational feature extraction algorithms.

The results indicated that the difference between the measurements (automated measurement - expert manual measurement) for the pre-op extrameatal tumour measurement was very low (95\% agreement within -1.294 $mm$ to 2.848 $mm$). However, the difference in whole tumour measurement for intrameatal only sessions was higher with 95\% agreement between -3.277 $mm$ and 3.138 $mm$.

Our findings indicate that the automated features currently overestimate linear measurements for pre-operative extrameatal region and post-operative intrameatal region. Over-segmentation of structures using the deep learning model might have caused this over-estimation. Further, features extracted by the expert were performed using T1C modality if available (T2 only if T1C was not available). This methodology differed slightly to our automated framework where the T2 modality was used preferentially if both modalities were available. This methodology may have resulted in biases in the automated measurements.

%\section{Conclusion}
In conclusion, our framework generates a clinical usable features that can guide the Vestibular Schwannoma decision making. Our results indicate a promising expansion of this work. These features could be incorporated into a formal report that may be used support clinical decision making by the specialist multidisciplinary team meetings. This work could also be adapted to other benign brain tumours to output tumour-specific features.

\ifdefined\showidentifiedinfo
\section{Disclosures}
N. Wijethilake was supported by the UK Medical Research Council [MR/N013700/1] and the King’s College London MRC Doctoral Training Partnership in Biomedical Sciences. This work was supported by Wellcome Trust (203145Z/16/Z, 203148/Z/16/Z, WT106882), EPSRC (NS/A000050/1, NS/A000049/1) and MRC (MC/PC/180520) funding. Additional funding was provided by Medtronic. TV is also supported by a Medtronic/Royal Academy of Engineering Research Chair (RCSRF1819/7/34). SO is co-founder and shareholder of BrainMiner Ltd, UK.
\fi
% References
\bibliography{report} % bibliography data in report.bib

\begin{thebibliography}{10}

\bibitem{shapey2018standardised}
Shapey, J., Barkas, K., Connor, S., Hitchings, A., Cheetham, H., Thomson, S., U-King-Im, J., Beaney, R., Jiang, D., Barazi, S., et~al., ``A standardised pathway for the surveillance of stable vestibular schwannoma,'' {\em The Annals of The Royal College of Surgeons of England}~{\bf 100}(3),  216--220 (2018).

\bibitem{prasad2018decision}
Prasad, S.~C., Patnaik, U., Grinblat, G., Giannuzzi, A., Piccirillo, E., Taibah, A., and Sanna, M., ``Decision making in the wait-and-scan approach for vestibular schwannomas: is there a price to pay in terms of hearing, facial nerve, and overall outcomes?,'' {\em Neurosurgery}~{\bf 83}(5),  858--870 (2018).

\bibitem{kanzaki2003new}
Kanzaki, J., Tos, M., Sanna, M., and Moffat, D.~A., ``New and modified reporting systems from the consensus meeting on systems for reporting results in vestibular schwannoma,'' {\em Otology \& neurotology}~{\bf 24}(4),  642--649 (2003).

\bibitem{pyradiomics}
Van~Griethuysen, J.~J., Fedorov, A., Parmar, C., Hosny, A., Aucoin, N., Narayan, V., Beets-Tan, R.~G., Fillion-Robin, J.-C., Pieper, S., and Aerts, H.~J., ``Computational radiomics system to decode the radiographic phenotype,'' {\em Cancer research}~{\bf 77}(21),  e104--e107 (2017).

\bibitem{nioche2018lifex}
Nioche, C., Orlhac, F., Boughdad, S., Reuz{\'e}, S., Goya-Outi, J., Robert, C., Pellot-Barakat, C., Soussan, M., Frouin, F., and Buvat, I., ``Lifex: a freeware for radiomic feature calculation in multimodality imaging to accelerate advances in the characterization of tumor heterogeneity,'' {\em Cancer research}~{\bf 78}(16),  4786--4789 (2018).

\bibitem{hatt2018ibsi}
Hatt, M., Vallieres, M., Visvikis, D., and Zwanenburg, A., ``Ibsi: an international community radiomics standardization initiative,'' (2018).

\bibitem{mackeith2018comparison}
MacKeith, S., Das, T., Graves, M., Patterson, A., Donnelly, N., Mannion, R., Axon, P., and Tysome, J., ``A comparison of semi-automated volumetric vs linear measurement of small vestibular schwannomas,'' {\em European Archives of Oto-Rhino-Laryngology}~{\bf 275}(4),  867--874 (2018).

\bibitem{shapey2021segmentation}
Shapey, J., Kujawa, A., Dorent, R., Wang, G., Dimitriadis, A., Grishchuk, D., Paddick, I., Kitchen, N., Bradford, R., Saeed, S.~R., et~al., ``Segmentation of vestibular schwannoma from mri, an open annotated dataset and baseline algorithm,'' {\em Scientific Data}~{\bf 8}(1),  286 (2021).

\bibitem{shapey2021artificial}
Shapey, J., Kujawa, A., Dorent, R., Saeed, S.~R., Kitchen, N., Obholzer, R., Ourselin, S., Vercauteren, T., and Thomas, N.~W., ``Artificial intelligence opportunities for vestibular schwannoma management using image segmentation and clinical decision tools,'' {\em World neurosurgery}~{\bf 149},  269--270 (2021).

\bibitem{aaronnmm}
Kujawa, A., Dorent, R., Connor, S., Thomson, S., Ivory, M., Vahedi, A., Guilhem, E., Bradford, R., Kitchen, N., Bisdas, S., et~al., ``Deep learning for automatic segmentation of vestibular schwannoma: A retrospective study from multi-centre routine mri,'' {\em medRxiv} ,  2022--08 (2022).

\bibitem{shapey2019artificial}
Shapey, J., Wang, G., Dorent, R., Dimitriadis, A., Li, W., Paddick, I., Kitchen, N., Bisdas, S., Saeed, S.~R., Ourselin, S., et~al., ``An artificial intelligence framework for automatic segmentation and volumetry of vestibular schwannomas from contrast-enhanced t1-weighted and high-resolution t2-weighted mri,'' {\em Journal of neurosurgery}~{\bf 134}(1),  171--179 (2019).

\bibitem{wijethilake2022boundary}
Wijethilake, N., Kujawa, A., Dorent, R., Asad, M., Oviedova, A., Vercauteren, T., and Shapey, J., ``Boundary distance loss for intra-/extra-meatal segmentation of vestibular schwannoma,'' in [{\em Machine Learning in Clinical Neuroimaging: 5th International Workshop, MLCN 2022, Held in Conjunction with MICCAI 2022, Singapore, September 18, 2022, Proceedings}{\nolinebreak\hspace{0.1em}]},   73--82, Springer (2022).

\bibitem{dorent2023crossmoda}
Dorent, R., Kujawa, A., Ivory, M., Bakas, S., Rieke, N., Joutard, S., Glocker, B., Cardoso, J., Modat, M., Batmanghelich, K., et~al., ``Crossmoda 2021 challenge: Benchmark of cross-modality domain adaptation techniques for vestibular schwannoma and cochlea segmentation,'' {\em Medical Image Analysis}~{\bf 83},  102628 (2023).

\bibitem{isensee2021nnu}
Isensee, F., Jaeger, P.~F., Kohl, S.~A., Petersen, J., and Maier-Hein, K.~H., ``nnu-net: a self-configuring method for deep learning-based biomedical image segmentation,'' {\em Nature methods}~{\bf 18}(2),  203--211 (2021).

\bibitem{andrew1979another}
Andrew, A.~M., ``Another efficient algorithm for convex hulls in two dimensions,'' {\em Information Processing Letters}~{\bf 9}(5),  216--219 (1979).

\end{thebibliography}
\bibliographystyle{spiebib} % makes bibtex use spiebib.bst

\end{document}